\documentclass[nonacm,sigchi]{acmart}

\renewcommand\footnotetextcopyrightpermission[1]{}
\usepackage{amsmath}
\usepackage{amssymb}

\usepackage{threeparttable}
\usepackage{array}
\usepackage{multirow}

\AtBeginDocument{%
  \providecommand\BibTeX{{%
    \normalfont B\kern-0.5em{\scshape i\kern-0.25em b}\kern-0.8em\TeX}}}

\begin{document}

\title[EmbraceNet for Activity: A Deep Multimodal Fusion Architecture for Activity Recognition]{EmbraceNet for Activity: A Deep Multimodal Fusion Architecture for Activity Recognition}
\thanks{* Accepted in International Workshop on Human Activity Sensing Corpus and Applications (HASCA) 2019 at ACM International Joint Conference on Pervasive and Ubiquitous Computing and International Symposium on Wearable Computers (UbiComp/ISWC)}

\author{Jun-Ho Choi}
\affiliation{%
  \institution{School of Integrated Technology, Yonsei University}
  \city{Incheon}
  \country{Korea}
}
\email{idearibosome@yonsei.ac.kr}
\email{}
\email{}

\author{Jong-Seok Lee}
\affiliation{%
	\institution{School of Integrated Technology, Yonsei University}
	\city{Incheon}
	\country{Korea}
}
\email{jong-seok.lee@yonsei.ac.kr}
\email{}
\email{}

\renewcommand{\shortauthors}{Choi and Lee}

\begin{abstract}
Human activity recognition using multiple sensors is a challenging but promising task in recent decades.
In this paper, we propose a deep multimodal fusion model for activity recognition based on the recently proposed feature fusion architecture named EmbraceNet.
Our model processes each sensor data independently, combines the features with the EmbraceNet architecture, and post-processes the fused feature to predict the activity.
In addition, we propose additional processes to boost the performance of our model.
We submit the results obtained from our proposed model to the SHL recognition challenge with the team name ``Yonsei-MCML.''
\end{abstract}


\maketitle

\section{Introduction}

Human activity recognition is one of the widely studied topics in the recent decades \cite{herath2017going}.
This has been accelerated with the abundance of computational resources and data.
Thanks to the popularization of the intensive computing devices such as graphics processing units (GPUs) \cite{abadi2016tensorflow}, it becomes possible to build large deep neural network models containing considerable numbers of the layers \cite{simonyan2014very}.
Furthermore, several datasets for the activity recognition tasks containing various types of data have been released, including videos (e.g., HMDB \cite{kuehne2011hmdb} and ActivityNet \cite{caba2015activitynet}), inertial sensors (e.g., OPPORTUNITY \cite{chavarriaga2013opportunity} and Sussex-Huawei Locomotion-
Transportation (SHL) \cite{gjoreski2018university,wang2019enabling}), and multiple types of the sensors (e.g., Berkeley MHAD \cite{ofli2013berkeley} and UTD-MHAD \cite{chen2015utd}).

Along with the increased number of sensors, many modality fusion methods have been proposed to efficiently handle the characteristics of the multimodal data \cite{ramachandram2017deep}.
For instance, Ord{\'o}{\~n}ez and Roggen \cite{ordonez2016deep} employed an early fusion-based approach, which integrates the sensor data at the initial stage, to classify actions of the OPPORTUNITY dataset.
Ngiam \textit{et al.} \cite{ngiam2011multimodal} proposed an intermediate fusion-based approach to exploit audio and visual features of videos, which uses a restricted Boltzmann machine (RBM) for each modality to obtain the features and combines them at the intermediate stage.

In this paper, we propose an intermediate fusion-based deep activity recognition model, which pre-processes the data of each modality separately and then combines them with a state-of-the-art multimodal fusion method named as EmbraceNet \cite{choi2019embracenet}.
EmbraceNet brings several advantages to the multimodal classification tasks, including high compatibility with existing deep learning architectures and thorough consideration of the cross-modal correlations.
Furthermore, we employ additional processes such as data pre-processing, data augmentation, and iterative application of our model to improve the performance.
We participate in the SHL recognition challenge, which employs part of the SHL dataset \cite{gjoreski2018university,wang2019enabling}, under the team name of ``Yonsei-MCML.''

The rest of the paper is organized as follows.
First, we explain the dataset that we use in Section~\ref{sec:dataset}.
Then, we present the structure of our activity recognition model in Section~\ref{sec:model_description}.
We conduct several experiments to analyze our proposed model in Section~\ref{sec:experiments}.
Finally, we conclude our work in Section~\ref{sec:conclusion} with providing additional information about our submission to the SHL recognition challenge.

\begin{figure*}
	\begin{center}
		\centering
		\includegraphics[width=0.99\linewidth]{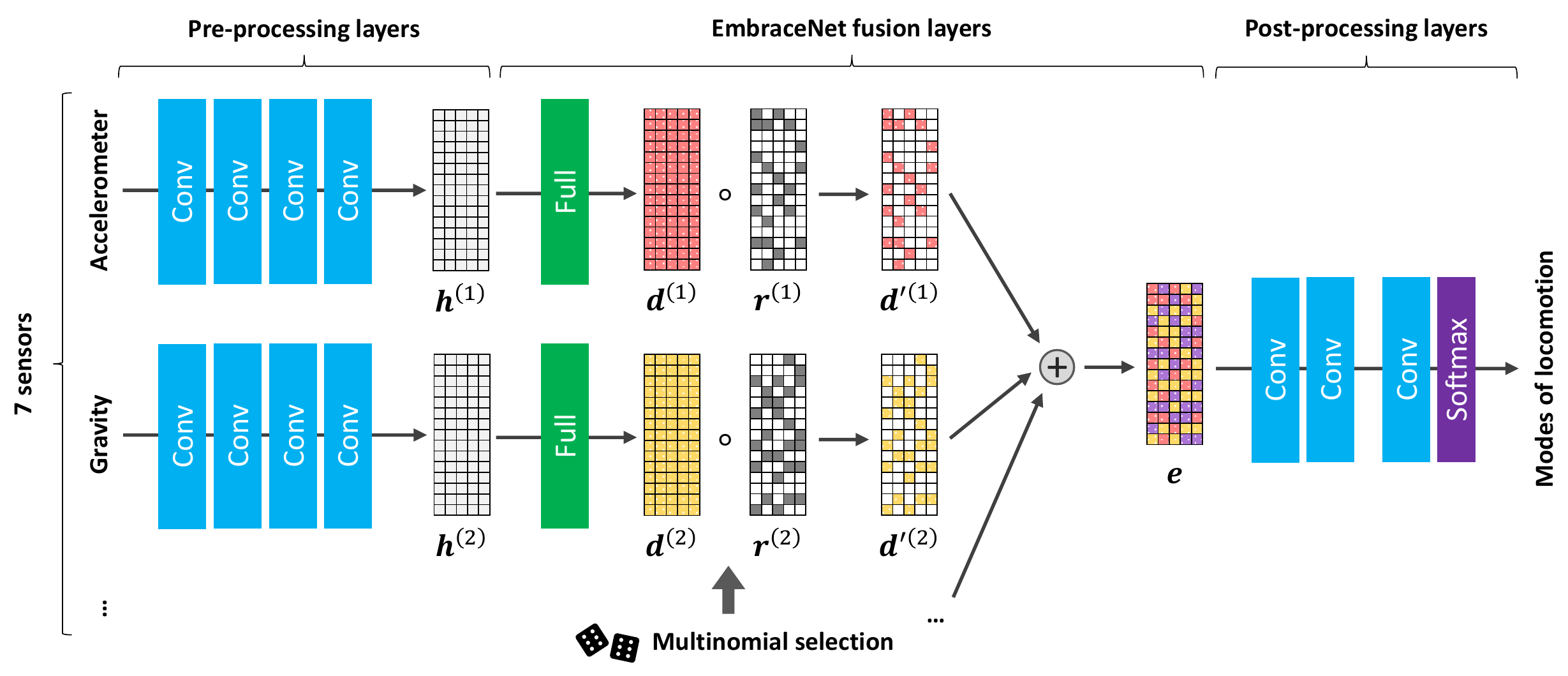}
	\end{center}
	\caption{Overall structure of our activity recognition model.}
	\label{fig:base_model_structure}
\end{figure*}

\begin{table}
	\small
	\caption{Model configuration of our activity recognition model.}
	\label{table:base_model_config}
	\begin{tabular}{c|c|c|c}
		\toprule
		\textbf{Part} & \textbf{Layer} & \textbf{Configuration} & \textbf{Output size} \\
		\midrule
		\multirow{4}{*}{Pre-processing} & conv1 & 32 filters, 5 strides & $100 \times 32$ \\
		& conv2 & 64 filters, 5 strides & $20 \times 64$ \\
		& conv3 & 128 filters, 2 strides & $10 \times 128$ \\
		& conv4 & 256 filters, 2 strides & $5 \times 256$ \\
		\midrule
		EmbraceNet & full & 256 units & $5 \times 256$ \\
		\midrule
		\multirow{3}{*}{Post-processing} & conv1 & 256 filters, 1 stride & $5 \times 256$ \\
		& conv2 & 256 filters, 1 stride & $5 \times 256$ \\
		& conv3 & 8 filters, 1 stride & $5 \times 8$ \\
		\bottomrule
	\end{tabular}
\end{table}

\section{Dataset}
\label{sec:dataset}

The SHL dataset contains eight modes of locomotion and transportation, including \textit{still}, \textit{walking}, \textit{run}, \textit{bike}, \textit{car}, \textit{bus}, \textit{train}, and \textit{subway}.
The sensor data are acquired using smartphones with four locations of a person, including bag, trousers front pocket, breast pocket, and a hand.
Each location is referred to as ``Bag'', ``Hips'', ``Torso'', and ``Hand'', respectively.

A subset of the dataset is provided for the challenge, which consists of the data obtained from seven sensors, including accelerometer, gravity, gyroscope, linear accelerometer, magnetometer, orientation, and pressure sensors.
The orientation sensor consists of four channels, which is represented as spatial rotation in quaternions (i.e., $w$, $x$, $y$, and $z$-axes).
The pressure sensor consists of a single channel.
Each of the rest sensors consists of three channels, which correspond to the three-dimensional axes (i.e., $x$, $y$, and $z$-axes).

The dataset is provided with three splits: train, validation, and test.
For training, 59 days of the data acquired from the locations except the ``Hand'' location are provided.
For validation, 3 days of the data obtained from all the locations are given.
For testing, 20 days of the data for only the ``Hand'' location are provided.
Both the training and validation splits contain the ground-truth labels along with the sensor data, whereas the testing split does not.

The three splits consist of 588,215, 48,708, and 55,811 samples, respectively\footnote{The original number of the training samples is 588,216, however, we exclude the sample that contains invalid data ($\mathrm{NaN}$).}.
Each sample contains 500 sensor values, which are acquired for five seconds with a sampling rate of 100 Hz.

\section{Model Description}
\label{sec:model_description}

We develop a multimodal activity recognition model, which processes each modality data independently in pre-processing layers, combines the processed features into one based on the EmbraceNet architecture \cite{choi2019embracenet}, and finally determines the activity classes via post-processing layers.
\figurename~\ref{fig:base_model_structure} shows the overall structure of our model.
The detailed model configuration is explained in Table~\ref{table:base_model_config}.

\subsection{Pre-processing layers}

We employ all the $m=7$ sensors to recognize the mode of locomotion or transportation.
Denote the original data of the $k$-th sensor as $\mathbf{x}^{(k)}$.
The shape of $\mathbf{x}^{(k)}$ is ${t}\times{s}$, where the first and second dimensions (i.e., $t$ and $s$) correspond to time and channel axes, respectively.
For example, the accelerometer sensor data $\mathbf{x}^{(1)}$ has a shape of $500 \times 3$.
For each sensor, the original data is independently processed by the one-dimensional convolutional layers having a kernel size of $5$ and the ReLU activation (i.e., $R(x) = \max{(0, x)}$).
In addition, each convolutional layer employs convolutional operations with strides, which reduces the size of the output along the time axis.
The pre-processing layers finally output $m$ processed features $\{\mathbf{h}^{(1)}, \mathbf{h}^{(2)}, ..., \mathbf{h}^{(m)}\}$, each of which has a shape of $5\times256$.

\subsection{EmbraceNet fusion layers}

For each pre-processed feature $\mathbf{h}^{(k)}$, our model first employs a fully-connected layer having $c=256$ units with the ReLU activation, which produces another processed feature $\mathbf{d}^{(k)}$.
Then, the EmbraceNet-based feature fusion \cite{choi2019embracenet} proceeds as follows.
Let $\mathbf{p} = [{p}_{1}, {p}_{2}, ..., {p}_{m}]^{T}$ denote a vector that consists of probability values (i.e., $\sum_{k}{p_k} = 1$).
Then, a vector $\mathbf{r}_{ij} = [{r}_{ij}^{(1)}, {r}_{ij}^{(2)}, ..., {r}_{ij}^{(m)}]^{T}$ can be drawn from a multinomial distribution, i.e.,
\begin{equation}
	\label{eq:embracenet_toggle_multinomial}
	\mathbf{r}_{ij} \sim \textrm{Multinomial}(1, \mathbf{p})
\end{equation}
where $i \in \{1, 2, ..., 5\}$ and $j \in \{1, 2, ..., c\}$.
Note that this ensures only one value of $\mathbf{r}_{ij}$ is 1 and the rest values are 0.
Then, a feature mask $\textbf{r}^{(k)} \in \mathbb{R}^{{5}\times{c}}$ can be obtained, where the value of $\textbf{r}^{(k)}$ at $(i, j)$ is ${r}_{ij}^{(k)}$.
This is applied to $\mathbf{d}^{(k)}$ as
\begin{equation}
	\label{eq:embracenet_masking}
	\mathbf{d}^{\prime(k)} = \mathbf{r}^{(k)} \circ \mathbf{d}^{(k)}
\end{equation}
where $\circ$ denotes the Hadamard product (i.e., ${d}_{ij}^{\prime(k)} = {r}_{ij}^{(k)} \cdot {d}_{ij}^{(k)}$).
Finally, our model combines all the masked features to obtain a final fused representation $\mathbf{e}$, i.e.,
\begin{equation}
	\label{eq:embracenet_final}
	\mathbf{e} = \sum_{k}{\mathbf{d}^{\prime(k)}}.
\end{equation}
In this work, we set $\mathbf{p} = [1/m, 1/m, ..., 1/m]^{T}$.


\subsection{Post-processing layers}

The fused representation is further processed in the post-processing layers, which consist of additional one-dimensional convolutional layers having a kernel size of $5$ and the ReLU activation.
Then, the final convolutional layer (conv3 in Table~\ref{table:base_model_config}) consisting of 8 filters with a kernel size of $1$ and a softmax activation is used to obtain the final class probabilities, among which the maximum is chosen as the recognized class.
A total of five decisions are obtained, each of which corresponds to the result for each one-second time duration.

\section{Experiments}
\label{sec:experiments}

We conduct experiments to compare the performance of our method with the other fusion methods and find ways to improve the performance of our method.
All the results are evaluated on the validation split.

\subsection{Implementation details}

We implement our model on the TensorFlow framework \cite{abadi2016tensorflow}.
For each training step, eight samples are randomly chosen from the train split.
We employ a cross entropy loss and the Adam optimization method \cite{kingma2014adam} with ${\beta}_{1} = 0.9$, ${\beta}_{2} = 0.999$, and $\hat{\epsilon} = {10}^{-4}$.
The learning rate is initially set to ${10}^{-4}$ and reduced by a factor of 2 at every $1 \times {10}^{5}$ steps.
A total of $5 \times {10}^{5}$ training steps are executed.

\begin{table}
	\small
	\caption{Performance comparison of the modality fusion methods.}
	\label{table:performance_fusion_methods}
	\begin{tabular}{l|c}
		\toprule
		\textbf{Fusion method} & \textbf{Accuracy} \\
		\midrule
		Early fusion & 46.73\% \\
		Intermediate fusion & 63.87\% \\
		Late fusion & 63.85\% \\
		Confidence fusion \cite{choi2018confidence} & 63.60\% \\
		Ours & \textbf{65.22\%} \\
		\bottomrule
	\end{tabular}
\end{table}

\begin{table}
	\small
	\caption{Performance comparison of our model trained with data obtained from different phone locations.}
	\label{table:performance_phone_locations}
	\begin{tabular}{l|c}
		\toprule
		\textbf{Trained location} & \textbf{Accuracy (Hand)} \\
		\midrule
		Bag & 30.84\% \\
		Hips & 36.32\% \\
		Torso & 30.87\% \\
		All & \textbf{47.63\%} \\
		\bottomrule
	\end{tabular}
\end{table}

\begin{table*}
	\small
	\caption{Performance comparison of our model trained on the processed inputs.}
	\label{table:performance_input_processing}
	\begin{tabular}{lc|ccccc}
		\toprule
		\textbf{Input} & \textbf{Random rotation} & \textbf{Accuracy (Bag)} & \textbf{Accuracy (Hips)} & \textbf{Accuracy (Torso)} & \textbf{Accuracy (Hand)} & \textbf{Accuracy (All)} \\
		\midrule
		Raw & No & 63.68\% & 67.98\% & 81.58\% & 47.63\% & 65.22\% \\
		FFT & No & 84.29\% & \textbf{84.55\%} & \textbf{87.21\%} & 54.28\% & 77.58\% \\
		Raw \& FFT & No & 78.85\% & 79.82\% & 86.74\% & 52.60\% & 74.50\% \\
		Raw & Yes & 81.55\% & 73.53\% & 77.91\% & \textbf{60.06\%} & 73.26\% \\
		FFT & Yes & 84.54\% & 81.55\% & 82.86\% & 58.11\% & 76.77\% \\
		Raw \& FFT & Yes & \textbf{85.35\%} & 82.21\% & 83.48\% & 59.34\% & \textbf{77.60\%} \\
		\bottomrule
	\end{tabular}
\end{table*}

\subsection{Comparison with the other fusion methods}

We first compare our model with the other fusion models: early, intermediate, late, and confidence \cite{choi2018confidence} fusion.
All the compared models have the same model configuration as our model (Table~\ref{table:base_model_config}) except the EmbraceNet part.
In the early fusion model, all the sensor data are concatenated along their channel dimension and then fed to the model.
In the intermediate fusion model, the EmbraceNet fusion layers are replaced with the concatenation of $\mathbf{h}^{(k)}$ along their last dimension.
In the late fusion model, an independent model is trained for each sensor and then the decision is made from the averaged softmax outputs.
In the confidence fusion model, the EmbraceNet fusion layers are replaced with the confidence calculation and fusion layers explained in \cite{choi2018confidence}.

Table~\ref{table:performance_fusion_methods} shows the performance comparison in terms of accuracy measured on the data from all phone locations.
The result confirms that our model outperforms the other methods.
The early fusion model shows the lowest accuracy.
This implies that combining sensor features at the intermediate or late stages is more beneficial than combining the data at the early stage.

\subsection{Training on different phone locations}

The train split does not contain any data acquired from the ``Hand'' location.
Therefore, classifying the ``Hand'' data in the validation (or test) split is a challenging situation due to the mismatch between the training and test data.
To examine the effect of this, we train our model with only one of the three phone locations (i.e., ``Bag'', ``Hips'', and ``Torso'') or all of them.
Table~\ref{table:performance_phone_locations} shows the performance in terms of accuracy measured on the ``Hand'' location.
The result shows that the ``Hips'' location is the most beneficial to classify the data from the ``Hand'' location.
Nevertheless, employing all the locations is much better than employing only one specific phone location.

\begin{table*}
	\small
	\caption{Performance comparison of our model with the output self-ensemble.}
	\label{table:performance_output_ensemble}
	\begin{tabular}{c|ccccc}
		\toprule
		\textbf{Self-ensemble size} & \textbf{Accuracy (Bag)} & \textbf{Accuracy (Hips)} & \textbf{Accuracy (Torso)} & \textbf{Accuracy (Hand)} & \textbf{Accuracy (All)} \\
		\midrule
		1 & 85.35\% & 82.21\% & 83.48\% & 59.34\% & 77.60\% \\
		2 & 86.08\% & 82.83\% & 84.24\% & 59.50\% & 78.16\% \\
		3 & 86.17\% & 83.09\% & 84.54\% & 60.19\% & 78.50\% \\
		4 & 86.31\% & \textbf{83.21\%} & 84.53\% & 60.03\% & 78.52\% \\
		5 & \textbf{86.35\%} & 83.11\% & \textbf{84.71\%} & \textbf{60.30\%} & \textbf{78.62\%} \\
		\bottomrule
	\end{tabular}
\end{table*}

\subsection{Data pre-processing}

It may be helpful to pre-process the raw data before inputting them to the model for performance improvement.
To examine this, we consider two data processing methods as follows.

\textbf{\textit{Fourier transform.}}
It is known that manipulating inertial sensor data in the frequency domain is beneficial to improve the classification performance \cite{gjoreski2018applying,ito2018application}.
To adopt this, we apply the one-dimensional fast Fourier transform (FFT) for each sensor channel before feeding the data to the model.

\textbf{\textit{Random rotation.}}
The inertial sensors output the data depending on the absolute orientation of the phone.
To make sure that the model is not biased towards the absolute orientation of the phone, we employ a random rotation process that changes the absolute orientation of the input data.
For a given inertial sensor data sample, three random rotation angles are drawn, which correspond to $x$, $y$, and $z$-axes, respectively.
Then, all values of the given data sample are manipulated so as to have the absolute orientation that is rotated by the randomly drawn angles.
We apply the random rotation process for all the sensor data except the pressure sensor.
For the orientation sensor data, we first convert the quaternions to the Euler angles (i.e., yaw, pitch, and roll), apply the random rotation, and convert them back to the quaternions.

Table~\ref{table:performance_input_processing} compares the performance of our models trained with different inputs.
A total of 6 models are obtained with the combination of the original (raw) data, data manipulated via FFT, and the random rotation process.
First of all, both the Fourier transform and random rotation contribute to improve the performance of our model.
For example, employing FFT significantly increases the accuracies measured on ``Hips'' and ``Torso'', and employing the random rotation largely improves the performance on the ``Hand'' location.
The model trained with employing both pre-processing methods achieves the highest accuracy when all validation data are used.
These results confirm that our data processing methods are highly beneficial to improve the performance.

\subsection{Output self-ensemble}

In the EmbraceNet fusion layers, the data flow to build an integrated representation (i.e., $\mathbf{e}$) is not static due to the stochastic multinomial sampling process.
In other words, our model may output different results at different inferences, even if the same input is given.
This implies that combining the outputs obtained by employing our model multiple times can have an effect similar to the \textit{ensemble learning} (or \textit{bootstrap aggregating}), which combines the decisions of multiple classifiers to boost the performance.

To investigate this opportunity, we employ the trained model multiple times, combine the outputs (after the softmax function), and make the decision.
Table~\ref{table:performance_output_ensemble} shows the performance of our model with respect to the number of the employing times.
This result strongly supports that the so-called \textit{output self-ensemble} process can improve the performance without any additional training process.

\subsection{Increasing model size}
\label{sec:increasing_model_size}

\begin{table}
	\small
	\caption{Model configuration of our activity recognition model having increased size.}
	\label{table:final_model_config}
	\begin{tabular}{c|c|c|c}
		\toprule
		\textbf{Part} & \textbf{Layer} & \textbf{Configuration} & \textbf{Output size} \\
		\midrule
		\multirow{10}{*}{Pre-processing} & conv1 & 32 filters, 1 stride & $500 \times 32$ \\
		& conv2 & 32 filters, 5 strides & $100 \times 32$ \\
		& conv3 & 64 filters, 1 stride & $100 \times 64$ \\
		& conv4 & 64 filters, 5 strides & $20 \times 64$ \\
		& conv5 & 128 filters, 1 stride & $20 \times 128$ \\
		& conv6 & 128 filters, 2 strides & $10 \times 128$ \\
		& conv7 & 256 filters, 1 stride & $10 \times 256$ \\
		& conv8 & 256 filters, 2 strides & $5 \times 256$ \\
		& conv9 & 512 filters, 1 stride & $5 \times 512$ \\
		& conv10 & 512 filters, 1 stride & $5 \times 512$ \\
		\midrule
		EmbraceNet & full & 512 units & $5 \times 512$ \\
		\midrule
		\multirow{5}{*}{Post-processing} & conv1 & 512 filters, 1 stride & $5 \times 512$ \\
		& conv2 & 512 filters, 1 stride & $5 \times 512$ \\
		& conv3 & 256 filters, 1 stride & $5 \times 256$ \\
		& conv4 & 256 filters, 1 stride & $5 \times 256$ \\
		& conv5 & 8 filters, 1 stride & $5 \times 8$ \\
		\bottomrule
	\end{tabular}
\end{table}

Finally, we increase the model size to enable more thorough analysis of the input features in our model as shown in Table~\ref{table:final_model_config}.
In detail, the number of the convolutional layers in the pre-processing layers increases from 4 to 10, the number of hidden units in the EmbraceNet fusion layers (i.e., $c$) increases from 256 to 512, and the number of the convolutional layers in the post-processing layers increases from 3 to 5.
With this model, we achieve improved accuracies of 87.10\%, 85.57\%, 86.66\%, and 63.03\% on the ``Bag'', ``Hips'', ``Torso'', and ``Hand'' locations, respectively, and 80.59\% on all the validation data.

\section{Conclusion}
\label{sec:conclusion}

In this paper, we proposed a deep learning-based multimodal activity recognition model that utilizes multiple sensor data efficiently.
Our model employed the pre-processing layers to analyze each sensor data, EmbraceNet-based fusion layers to combine the pre-processed features to an integrated representation, and the post-processing layers to make a decision.
We showed that our model outperformed the other fusion methods.

We submitted the results obtained on the test split to the SHL recognition challenge.
To do this, we trained the model explained in Table~\ref{table:final_model_config} on both the train and validation splits.
The file size of the trained (frozen) model is 167MB.
Both the Fourier transform and random rotation processes were employed as the data pre-processing methods.
The model was trained for ${5}\times{10}^{5}$ steps on a NVIDIA GeForce GTX 1080 GPU, which took about 17 hours.
A total of five activity probabilities were obtained for each sample to employ the output self-ensemble method.
It took about 25 minutes to obtain the results for all test samples.
The recognition result for the testing dataset will be presented in the summary paper of the challenge \cite{wang2019summary}.

\begin{acks}
This research was supported by the MSIT (Ministry of Science and ICT), Korea, under the ``ICT Consilience Creative Program'' (IITP-2019-2017-0-01015) supervised by the IITP (Institute for Information \& Communications Technology Planning \& Evaluation). In addition, this work was also supported by the IITP grant funded by the Korea government (MSIT) (R7124-16-0004, Development of Intelligent Interaction Technology Based on Context Awareness and Human Intention Understanding).
\end{acks}

\bibliographystyle{ACM-Reference-Format}
\bibliography{hasca2019}


\begin{thebibliography}{19}


\ifx \showCODEN    \undefined \def \showCODEN     #1{\unskip}     \fi
\ifx \showDOI      \undefined \def \showDOI       #1{#1}\fi
\ifx \showISBNx    \undefined \def \showISBNx     #1{\unskip}     \fi
\ifx \showISBNxiii \undefined \def \showISBNxiii  #1{\unskip}     \fi
\ifx \showISSN     \undefined \def \showISSN      #1{\unskip}     \fi
\ifx \showLCCN     \undefined \def \showLCCN      #1{\unskip}     \fi
\ifx \shownote     \undefined \def \shownote      #1{#1}          \fi
\ifx \showarticletitle \undefined \def \showarticletitle #1{#1}   \fi
\ifx \showURL      \undefined \def \showURL       {\relax}        \fi
\providecommand\bibfield[2]{#2}
\providecommand\bibinfo[2]{#2}
\providecommand\natexlab[1]{#1}
\providecommand\showeprint[2][]{arXiv:#2}

\bibitem[\protect\citeauthoryear{Abadi, Barham, Chen, Chen, Davis, Dean, Devin,
  Ghemawat, Irving, Isard, et~al\mbox{.}}{Abadi et~al\mbox{.}}{2016}]%
        {abadi2016tensorflow}
\bibfield{author}{\bibinfo{person}{Mart{\'\i}n Abadi}, \bibinfo{person}{Paul
  Barham}, \bibinfo{person}{Jianmin Chen}, \bibinfo{person}{Zhifeng Chen},
  \bibinfo{person}{Andy Davis}, \bibinfo{person}{Jeffrey Dean},
  \bibinfo{person}{Matthieu Devin}, \bibinfo{person}{Sanjay Ghemawat},
  \bibinfo{person}{Geoffrey Irving}, \bibinfo{person}{Michael Isard},
  {et~al\mbox{.}}} \bibinfo{year}{2016}\natexlab{}.
\newblock \showarticletitle{{TensorFlow}: {A} system for large-scale machine
  learning}. In \bibinfo{booktitle}{\emph{Proceedings of the USENIX Symposium
  on Operating Systems Design and Implementation}}. \bibinfo{pages}{265--283}.
\newblock


\bibitem[\protect\citeauthoryear{Caba~Heilbron, Escorcia, Ghanem, and
  Carlos~Niebles}{Caba~Heilbron et~al\mbox{.}}{2015}]%
        {caba2015activitynet}
\bibfield{author}{\bibinfo{person}{Fabian Caba~Heilbron},
  \bibinfo{person}{Victor Escorcia}, \bibinfo{person}{Bernard Ghanem}, {and}
  \bibinfo{person}{Juan Carlos~Niebles}.} \bibinfo{year}{2015}\natexlab{}.
\newblock \showarticletitle{Activitynet: A large-scale video benchmark for
  human activity understanding}. In \bibinfo{booktitle}{\emph{Proceedings of
  the IEEE Conference on Computer Vision and Pattern Recognition}}.
  \bibinfo{pages}{961--970}.
\newblock


\bibitem[\protect\citeauthoryear{Chavarriaga, Sagha, Calatroni, Digumarti,
  Tr{\"o}ster, Mill{\'a}n, and Roggen}{Chavarriaga et~al\mbox{.}}{2013}]%
        {chavarriaga2013opportunity}
\bibfield{author}{\bibinfo{person}{Ricardo Chavarriaga}, \bibinfo{person}{Hesam
  Sagha}, \bibinfo{person}{Alberto Calatroni}, \bibinfo{person}{Sundara~Tejaswi
  Digumarti}, \bibinfo{person}{Gerhard Tr{\"o}ster}, \bibinfo{person}{Jos{\'e}
  del~R Mill{\'a}n}, {and} \bibinfo{person}{Daniel Roggen}.}
  \bibinfo{year}{2013}\natexlab{}.
\newblock \showarticletitle{The {O}pportunity challenge: {A} benchmark database
  for on-body sensor-based activity recognition}.
\newblock \bibinfo{journal}{\emph{Pattern Recognition Letters}}
  \bibinfo{volume}{34}, \bibinfo{number}{15} (\bibinfo{year}{2013}),
  \bibinfo{pages}{2033--2042}.
\newblock


\bibitem[\protect\citeauthoryear{Chen, Jafari, and Kehtarnavaz}{Chen
  et~al\mbox{.}}{2015}]%
        {chen2015utd}
\bibfield{author}{\bibinfo{person}{Chen Chen}, \bibinfo{person}{Roozbeh
  Jafari}, {and} \bibinfo{person}{Nasser Kehtarnavaz}.}
  \bibinfo{year}{2015}\natexlab{}.
\newblock \showarticletitle{{UTD-MHAD}: {A} multimodal dataset for human action
  recognition utilizing a depth camera and a wearable inertial sensor}. In
  \bibinfo{booktitle}{\emph{Proceedings of the IEEE International Conference on
  Image Processing}}. \bibinfo{pages}{168--172}.
\newblock


\bibitem[\protect\citeauthoryear{Choi and Lee}{Choi and Lee}{2018}]%
        {choi2018confidence}
\bibfield{author}{\bibinfo{person}{Jun-Ho Choi} {and}
  \bibinfo{person}{Jong-Seok Lee}.} \bibinfo{year}{2018}\natexlab{}.
\newblock \showarticletitle{Confidence-based deep multimodal fusion for
  activity recognition}. In \bibinfo{booktitle}{\emph{Proceedings of the ACM
  International Joint Conference and International Symposium on Pervasive and
  Ubiquitous Computing and Wearable Computers}}. \bibinfo{pages}{1548--1556}.
\newblock


\bibitem[\protect\citeauthoryear{Choi and Lee}{Choi and Lee}{2019}]%
        {choi2019embracenet}
\bibfield{author}{\bibinfo{person}{Jun-Ho Choi} {and}
  \bibinfo{person}{Jong-Seok Lee}.} \bibinfo{year}{2019}\natexlab{}.
\newblock \showarticletitle{{E}mbrace{N}et: {A} robust deep learning
  architecture for multimodal classification}.
\newblock \bibinfo{journal}{\emph{Information Fusion}}  \bibinfo{volume}{51}
  (\bibinfo{year}{2019}), \bibinfo{pages}{259--270}.
\newblock


\bibitem[\protect\citeauthoryear{Gjoreski, Ciliberto, Wang, Morales, Mekki,
  Valentin, and Roggen}{Gjoreski et~al\mbox{.}}{2018a}]%
        {gjoreski2018university}
\bibfield{author}{\bibinfo{person}{Hristijan Gjoreski},
  \bibinfo{person}{Mathias Ciliberto}, \bibinfo{person}{Lin Wang},
  \bibinfo{person}{Francisco Javier~Ordonez Morales}, \bibinfo{person}{Sami
  Mekki}, \bibinfo{person}{Stefan Valentin}, {and} \bibinfo{person}{Daniel
  Roggen}.} \bibinfo{year}{2018}\natexlab{a}.
\newblock \showarticletitle{The {U}niversity of {S}ussex-{H}uawei {L}ocomotion
  and {T}ransportation dataset for multimodal analytics with mobile devices}.
\newblock \bibinfo{journal}{\emph{IEEE Access}}  \bibinfo{volume}{6}
  (\bibinfo{year}{2018}), \bibinfo{pages}{42592--42604}.
\newblock


\bibitem[\protect\citeauthoryear{Gjoreski, Janko, Re{\v{s}}{\v{c}}i{\v{c}},
  Mlakar, Lu{\v{s}}trek, Bizjak, Slapni{\v{c}}ar, Marinko, Drobni{\v{c}}, and
  Gams}{Gjoreski et~al\mbox{.}}{2018b}]%
        {gjoreski2018applying}
\bibfield{author}{\bibinfo{person}{Martin Gjoreski}, \bibinfo{person}{Vito
  Janko}, \bibinfo{person}{Nina Re{\v{s}}{\v{c}}i{\v{c}}},
  \bibinfo{person}{Miha Mlakar}, \bibinfo{person}{Mitja Lu{\v{s}}trek},
  \bibinfo{person}{Jani Bizjak}, \bibinfo{person}{Ga{\v{s}}per
  Slapni{\v{c}}ar}, \bibinfo{person}{Matej Marinko}, \bibinfo{person}{Vid
  Drobni{\v{c}}}, {and} \bibinfo{person}{Matja{\v{z}} Gams}.}
  \bibinfo{year}{2018}\natexlab{b}.
\newblock \showarticletitle{Applying multiple knowledge to {S}ussex-{H}uawei
  locomotion challenge}. In \bibinfo{booktitle}{\emph{Proceedings of the ACM
  International Joint Conference and International Symposium on Pervasive and
  Ubiquitous Computing and Wearable Computers}}. \bibinfo{pages}{1488--1496}.
\newblock


\bibitem[\protect\citeauthoryear{Herath, Harandi, and Porikli}{Herath
  et~al\mbox{.}}{2017}]%
        {herath2017going}
\bibfield{author}{\bibinfo{person}{Samitha Herath}, \bibinfo{person}{Mehrtash
  Harandi}, {and} \bibinfo{person}{Fatih Porikli}.}
  \bibinfo{year}{2017}\natexlab{}.
\newblock \showarticletitle{Going deeper into action recognition: A survey}.
\newblock \bibinfo{journal}{\emph{Image and Vision Computing}}
  \bibinfo{volume}{60} (\bibinfo{year}{2017}), \bibinfo{pages}{4--21}.
\newblock


\bibitem[\protect\citeauthoryear{Ito, Cao, Shuzo, and Maeda}{Ito
  et~al\mbox{.}}{2018}]%
        {ito2018application}
\bibfield{author}{\bibinfo{person}{Chihiro Ito}, \bibinfo{person}{Xin Cao},
  \bibinfo{person}{Masaki Shuzo}, {and} \bibinfo{person}{Eisaku Maeda}.}
  \bibinfo{year}{2018}\natexlab{}.
\newblock \showarticletitle{Application of {CNN} for human activity recognition
  with {FFT} spectrogram of acceleration and gyro sensors}. In
  \bibinfo{booktitle}{\emph{Proceedings of the ACM International Joint
  Conference and International Symposium on Pervasive and Ubiquitous Computing
  and Wearable Computers}}. \bibinfo{pages}{1503--1510}.
\newblock


\bibitem[\protect\citeauthoryear{Kingma and Ba}{Kingma and Ba}{2015}]%
        {kingma2014adam}
\bibfield{author}{\bibinfo{person}{Diederik~P. Kingma} {and}
  \bibinfo{person}{Jimmy Ba}.} \bibinfo{year}{2015}\natexlab{}.
\newblock \showarticletitle{Adam: {A} method for stochastic optimization}. In
  \bibinfo{booktitle}{\emph{Proceedings of the International Conference on
  Learning Representations}}. \bibinfo{pages}{1--13}.
\newblock


\bibitem[\protect\citeauthoryear{Kuehne, Jhuang, Garrote, Poggio, and
  Serre}{Kuehne et~al\mbox{.}}{2011}]%
        {kuehne2011hmdb}
\bibfield{author}{\bibinfo{person}{Hildegard Kuehne}, \bibinfo{person}{Hueihan
  Jhuang}, \bibinfo{person}{Est{\'\i}baliz Garrote}, \bibinfo{person}{Tomaso
  Poggio}, {and} \bibinfo{person}{Thomas Serre}.}
  \bibinfo{year}{2011}\natexlab{}.
\newblock \showarticletitle{{HMDB}: {A} large video database for human motion
  recognition}. In \bibinfo{booktitle}{\emph{Proceedings of the IEEE
  International Conference on Computer Vision}}. \bibinfo{pages}{2556--2563}.
\newblock


\bibitem[\protect\citeauthoryear{Ngiam, Khosla, Kim, Nam, Lee, and Ng}{Ngiam
  et~al\mbox{.}}{2011}]%
        {ngiam2011multimodal}
\bibfield{author}{\bibinfo{person}{Jiquan Ngiam}, \bibinfo{person}{Aditya
  Khosla}, \bibinfo{person}{Mingyu Kim}, \bibinfo{person}{Juhan Nam},
  \bibinfo{person}{Honglak Lee}, {and} \bibinfo{person}{Andrew~Y Ng}.}
  \bibinfo{year}{2011}\natexlab{}.
\newblock \showarticletitle{Multimodal deep learning}. In
  \bibinfo{booktitle}{\emph{Proceedings of the International Conference on
  Machine Learning}}. \bibinfo{pages}{689--696}.
\newblock


\bibitem[\protect\citeauthoryear{Ofli, Chaudhry, Kurillo, Vidal, and
  Bajcsy}{Ofli et~al\mbox{.}}{2013}]%
        {ofli2013berkeley}
\bibfield{author}{\bibinfo{person}{Ferda Ofli}, \bibinfo{person}{Rizwan
  Chaudhry}, \bibinfo{person}{Gregorij Kurillo}, \bibinfo{person}{Ren{\'e}
  Vidal}, {and} \bibinfo{person}{Ruzena Bajcsy}.}
  \bibinfo{year}{2013}\natexlab{}.
\newblock \showarticletitle{Berkeley {MHAD}: {A} comprehensive multimodal human
  action database}. In \bibinfo{booktitle}{\emph{Proceedings of the IEEE
  Workshop on Applications of Computer Vision}}. \bibinfo{pages}{53--60}.
\newblock


\bibitem[\protect\citeauthoryear{Ord{\'o}{\~n}ez and Roggen}{Ord{\'o}{\~n}ez
  and Roggen}{2016}]%
        {ordonez2016deep}
\bibfield{author}{\bibinfo{person}{Francisco~Javier Ord{\'o}{\~n}ez} {and}
  \bibinfo{person}{Daniel Roggen}.} \bibinfo{year}{2016}\natexlab{}.
\newblock \showarticletitle{Deep convolutional and {LSTM} recurrent neural
  networks for multimodal wearable activity recognition}.
\newblock \bibinfo{journal}{\emph{Sensors}} \bibinfo{volume}{16},
  \bibinfo{number}{1} (\bibinfo{year}{2016}), \bibinfo{pages}{1--25}.
\newblock


\bibitem[\protect\citeauthoryear{Ramachandram and Taylor}{Ramachandram and
  Taylor}{2017}]%
        {ramachandram2017deep}
\bibfield{author}{\bibinfo{person}{Dhanesh Ramachandram} {and}
  \bibinfo{person}{Graham~W Taylor}.} \bibinfo{year}{2017}\natexlab{}.
\newblock \showarticletitle{Deep multimodal learning: {A} survey on recent
  advances and trends}.
\newblock \bibinfo{journal}{\emph{IEEE Signal Processing Magazine}}
  \bibinfo{volume}{34}, \bibinfo{number}{6} (\bibinfo{year}{2017}),
  \bibinfo{pages}{96--108}.
\newblock


\bibitem[\protect\citeauthoryear{Simonyan and Zisserman}{Simonyan and
  Zisserman}{2015}]%
        {simonyan2014very}
\bibfield{author}{\bibinfo{person}{Karen Simonyan} {and}
  \bibinfo{person}{Andrew Zisserman}.} \bibinfo{year}{2015}\natexlab{}.
\newblock \showarticletitle{Very deep convolutional networks for large-scale
  image recognition}. In \bibinfo{booktitle}{\emph{Proceedings of the
  International Conference on Learning Representations}}.
\newblock


\bibitem[\protect\citeauthoryear{Wang, Gjoreski, Ciliberto, Lago, Murao, Okita,
  and Roggen}{Wang et~al\mbox{.}}{2019a}]%
        {wang2019summary}
\bibfield{author}{\bibinfo{person}{Lin Wang}, \bibinfo{person}{Hristijan
  Gjoreski}, \bibinfo{person}{Mathias Ciliberto}, \bibinfo{person}{Paula Lago},
  \bibinfo{person}{Kazuya Murao}, \bibinfo{person}{Tsuyoshi Okita}, {and}
  \bibinfo{person}{Daniel Roggen}.} \bibinfo{year}{2019}\natexlab{a}.
\newblock \showarticletitle{Summary of the {S}ussex-{H}uawei
  locomotion-transportation recognition challenge 2019}. In
  \bibinfo{booktitle}{\emph{Proceedings of the ACM International Joint
  Conference and International Symposium on Pervasive and Ubiquitous Computing
  and Wearable Computers}}.
\newblock


\bibitem[\protect\citeauthoryear{Wang, Gjoreski, Ciliberto, Mekki, Valentin,
  and Roggen}{Wang et~al\mbox{.}}{2019b}]%
        {wang2019enabling}
\bibfield{author}{\bibinfo{person}{Lin Wang}, \bibinfo{person}{Hristijan
  Gjoreski}, \bibinfo{person}{Mathias Ciliberto}, \bibinfo{person}{Sami Mekki},
  \bibinfo{person}{Stefan Valentin}, {and} \bibinfo{person}{Daniel Roggen}.}
  \bibinfo{year}{2019}\natexlab{b}.
\newblock \showarticletitle{Enabling reproducible research in sensor-based
  transportation mode recognition with the {S}ussex-{H}uawei dataset}.
\newblock \bibinfo{journal}{\emph{IEEE Access}}  \bibinfo{volume}{7}
  (\bibinfo{year}{2019}), \bibinfo{pages}{10870--10891}.
\newblock


\end{thebibliography}

\end{document}